\documentstyle[12pt]{article}
\def\fract#1#2{ {\textstyle #1 \over \textstyle #2}}

\begin{document}

\begin{center}
{\large \bf RELATIVISTIC VIEW ON THE NUCLEAR EFFECTS IN THE 
DEEP INELASTIC SCATTERING ON DEUTRON} 
\vspace{13pt} \\
{\Large V.V.~Burov and A.V.~Molochkov}\\
\vspace{10pt}
{\sl BLTP, Joint Institute for Nuclear Research, 141980 Dubna,
Russia}
\end{center}

\vspace{10pt}
\begin{abstract}
We analyze nuclear effects in the deep inelastic scattering on deuteron
in the framework of the covariant approach.
It is shown that this approach gives us the way 
to investigate
the role of relativistic effects in the deep inelastic scattering (DIS)
on deuteron, such as the relativistic kinematics
and the off-mass-shell behavior of the Compton amplitude of nucleon
in the consistent manner.
We have obtained that taking into account of the nucleon amplitude
off-mass-shell behavior gives us an analog of the
interaction corrections in nonrelativistic models.
\end{abstract}

\section{ Introduction}

The cross section for the deep inelastic scattering
is proportional to the lepton and hadron tensors in the 
one photon approximation:
\begin{equation}
\sigma \propto L_{\mu\nu}(p,q)W^{\mu\nu}(P,q).
\end{equation}
Where P, p and q are the initial momenta of the target, 
lepton and virtual photon, respectively.
All information about target
and its nuclear properties is concentrated in the hadron tensor which
can be expressed in terms of the scalar structure functions:
\begin{eqnarray}
W_{\mu \nu}(P,q)=\left( -g_{\mu\nu} + \fract{q_\mu q_\nu}{q^2}\right)
F_1(x) +
\fract{1}{P\cdot q}\left(P_\mu - \fract{P\cdot q}{q^2}q_\mu\right)
\left(P_\nu - \fract{P\cdot q}{q^2}q_\nu\right)F_2(x). \label{lorentz}
\end{eqnarray}
In the Bjorken limit we can use the following relation:
$$\lim _{Q^2\rightarrow \infty }g_{\mu \nu }W^{\mu \nu }(P,q)
=-\fract 1x F_2(x).$$
Nuclear effects in the deep inelastic scattering appear as
a difference of the
structure functions of the bound and free nucleons.
This can be provided, for example,
by kinematics of bounded nucleons, off-mass-shell
effects, and by existence inside deuteron
of some additional degrees of freedom such as
nucleon-antinucleon pairs, exchange mesons, $\Delta$ isobar, 
and others.
Starting from the covariant approach \cite{form} based on 
the Bethe-Salpeter
formalism \cite{BS}, we can analyze
the role of relativistic kinematics and
off-mass-shell effects in a consistent manner.

This approach allows us to
write the impulse approximation for
the hadron tensor in the following form:
\begin{eqnarray}
\label{imp21}W_{\mu \nu }^{D}(P,q)=\int \fract{d^4k}{(2\pi )^4}
W_{\mu \nu
}^N\left(\fract P2+k,q\right)f^N(P,k) +
\int \fract{d^4k}{(2\pi )^4}W_{\mu \nu
}^{\overline{N}}\left(\fract P2+k,q\right)f^{\overline{N}}(P,k).
\label{hadron}
\end{eqnarray}
Where $k$ is the relative momentum
of nucleons inside deuteron.
The distribution functions $f^N(P,k)$ are expressed via the BS-vertex
 function
$\Gamma(P,k)$:
\begin{eqnarray}
f^N(P,k)=
-\overline{\Gamma}_{\alpha \beta }(P,k)S^{+}_{\alpha \gamma }
\left(\fract
P2+k\right)S_{\beta \delta }\left(\fract P2-k\right)
\Gamma_{\gamma \delta }(P,k)
\fract {1}{2E\left(\left(\fract{P}{2}+k\right)_0-E\right)},
\nonumber\\ \nonumber\\
f^{\overline N}(P,k)=
-\overline{\Gamma}_{\alpha \beta }(P,k)S^{-}_{\alpha \gamma }
\left(\fract
P2+k\right)S_{\beta \delta }\left(\fract P2-k\right)
\Gamma_{\gamma \delta }(P,k)
\fract {1}{2E\left(\left(\fract{P}{2}+k\right)_0+E\right)},\nonumber
\end{eqnarray}
$$E=\sqrt{\left({\bf \fract P2 + k}\right)^2+m^2}$$
Where $S(P,k)$ is a nucleon propagator and  $S^{+(-)}(P,k)$ 
is the propagator of
a nucleon with positive (negative) energy.
The letters $\alpha,\beta,\delta and \gamma$ denote the spinor indices.

The elementary nucleon (antinucleon) amplitude 
$W_{\mu \nu}^{N(\overline{N})}
\left(\fract P2+k,q\right)$
depends on the time component of the nucleon relative momentum, and
square of momentum of this nucleon is not equal to its mass $m$.
It leads to two difficulties at least. The first consists in changes of
the representation
(\ref{lorentz}) for this hadron tensor and the second is that we cannot use
the structure function of a physical nucleon to calculate 
a deuteron one.
To solve this situation, we have to express a deuteron hadron tensor in terms
of physical nucleon tensors.

\section{Expansion near Mass-Shell}

As one can see, expression (\ref{hadron}) is a convolution 
of the nucleon
hadron tensor and distribution function which contains information
about nuclear properties of the target. Inasmuch as deuteron is a 
weakly bounded
system we can try to expand this function near 
mass-shell of the nucleon
interacting with photon. 
To do it, one can rewrite nucleon propagator near
the pole as
\begin{equation}
\fract{1}{\left(\left(\fract P2 + k\right)_0-E\right)^2}\simeq
i\pi\delta^{\prime}
\left(\left(\fract P2 + k\right)_0-E\right).\label{pole}
\end{equation}
Substitution of this relation in the expression for the deuteron hadron tensor
gives us expansion of the latter in terms of 
the nucleon hadron tensor and its
derivatives at $\left(\fract P2 + k\right)_0=E$. First, we can use
the representation (\ref{lorentz}) to obtain the structure function. 
But the representation of a derivative 
of the on-mass-shell tensor consists 
of both the structure functions $F_1$, $F_2$ and the derivatives 
of additional
structure functions.
Neglecting derivatives of these additional structure functions, 
we can obtain
the following expression for the deuteron structure function 
in the laboratory system
(momentum of deuteron is $P=(M_D,{\bf 0})$,
momentum of photon is $q\simeq (q_0,0,0,q_0)$):
{\small
\begin{eqnarray}
F_2^D(x_D)=\int \fract{d^3k}{(2\pi)^3}
\fract{m^2}{4E^3(M_D-2E)^2}\left\{F_2^N(x_N)
\left(1-\fract{E+k_3}{M_D}\right)\Phi^2(M_D,k)
- \right. \label{f2}
\end{eqnarray}
\vspace*{-.5cm}
\begin{eqnarray}
\left. \fract{M_D-2E}{M_D}
x_N\fract{dF_2^N(x_N)}{dx_N}\Phi^2(M_D,k)
+ F_2^N(x_N)\fract{E-k_3}{M_D}(M_D-2E)
\fract{\partial}{\partial k_0}\Phi^2(M_D,k)\right\}_
{k_0=E-\fract{M_D}{2}}
\nonumber\end{eqnarray}
}
Here $x_D=\fract{-q^2}{M_Dq_0}$  and $x_N=\fract{-q^2}{(E-k_3)q_0}$
are Bjorken variables for the nucleon and deuteron, respectively.
The function $\Phi^2(M_D,k)$ is connected with the Bethe-Salpeter 
vertex
function $\Gamma(P,k)$ in the rest frame of the deuteron:
{\small
\begin{eqnarray}
\Phi^2(M_D,k)=\overline{\Gamma}_{\alpha\beta}(M_D,k)
\sum\limits_{s} u_\alpha^s({\bf k})\overline{u}_\delta^s({\bf k})
\sum\limits_{s^\prime}u_\beta^{s^\prime}(-{\bf k})
\overline{u}_\gamma^{s^\prime}(-{\bf k})\Gamma_{\delta\gamma}(M_D,k)
+\nonumber\\
\overline{\Gamma}_{\alpha\beta}(M_D,k)
\sum\limits_{s}v_\alpha^s(-{\bf k})\overline{v}_\delta^s(-{\bf k})
\sum\limits_{s^\prime}v_\beta^{s^\prime}({\bf k})
\overline{v}_\gamma^{s^\prime}({\bf k})\Gamma_{\delta\gamma}(M_D,k).
\end{eqnarray}
}
Here we neglect terms with  power of $\fract {<V>}{M_D}$ large than two;
$<V>$ is the mean value of the nucleon-nucleon potential.

\section{Nonrelativistic Limit}

To compare our result with the nonrelativistic calculations 
(for example
\cite{nonrel}), we expand $E$ in powers $\fract{{\bf p}^2}{m^2}$ in
(\ref{f2}) and discard the last term because of its pure relativistic
behavior.
That gives us the following expression for the structure function 
$F_2^D$:
\begin{eqnarray}
\fract{1}{2}F_2^D(x_D)=\int \fract{d^3k}{(2\pi)^3}
\left\{F_2^N(x_N)
\left(1-\fract{k_3}{m}+\fract{\epsilon}{m}\right)
\Psi^2({\bf k})
-
\right. \phantom{aaaaaaaaaaaaaaa}\label{f2nr}
\end{eqnarray}
\vspace*{-.5cm}
\begin{eqnarray}
\phantom{aaaaaaaaaaaaaaaaaaaaaaaaaaaaaa}\left.
\fract{-T+\epsilon}{m}
x_N\fract{dF_2^N(x_N)}{dx_N}\Psi^2({\bf k})\right\}
\nonumber\end{eqnarray}
$T=2E-2m$ is the kinetic energy of nucleons, $\epsilon=M-2m$ 
is the binding
energy.
Here we introduce an analog of the nonrelativistic wave 
function $\Psi^2({\bf k})$:
$$
\Psi^2({\bf k})=\fract{m^2}{4E^2M_D(M_D-2E)^2}
\left\{\Phi^2(M_D,k)\right\}_{k_0=E-\fract{M_D}{2}}
$$
with the usual normalization condition:
$$\int \fract{d^3k}{(2\pi)^3}\Psi^2({\bf k})=1.$$
Comparing (\ref{f2nr}) with nonrelativistic calculations we 
can conclude
that we have got here an analog of the nonrelativistic
impulse approximation with interaction corrections.

\section{Conclusion}
In the present talk we analyzed the relativistic impulse 
approximation for
the deuteron structure function $F_2^D$.
We droped out here some of the off-mass-shell effects connected with
an additional structure functions in (\ref{hadron}). This approximation
requires additional investigation. Also, we have not taken into account
the contribution from $P$-states to the Bethe-Salpeter vertex function. 
We will analyze its contribution in the near future.
We have obtained the contribution of effects resulting
from the Bethe-Salpeter
vertex function dependence of relative nucleon energy (\ref{f2}).
Com\-pa\-ring the obtained results with calculation in the nonrelativistic
field theory model \cite{nonrel} allows us to conclude
that the relativistic impulse approximation in the Bethe-Salpeter 
formalism
contains both the effects of the Fermi motion and some mesonic effects.
So relativistic calculation can shed light on nature of mesonic
corrections in the nonrelativistic models.

\end{document}